\documentclass[%
twocolumn,
 fleqn,usenatbib,
 showkeys,
 floatfix,nolongbibliography,aps,
author-numerical%
]{revtex4-2}

\pdfoutput=1
\usepackage[T1]{fontenc}
\usepackage [latin1]{inputenc}
\DeclareRobustCommand{\VAN}[3]{#2}
\let\VANthebibliography\thebibliography
\def\thebibliography{\DeclareRobustCommand{\VAN}[3]{##3}\VANthebibliography}

\usepackage{newtxtext,newtxmath}
\usepackage{tikz,xcolor,hyperref}
\usepackage[normalem]{ulem}
\usepackage{graphicx,url,caption,subcaption}
\usepackage{dcolumn}
\usepackage{bm}

\let\realhref\href

\definecolor{lime}{HTML}{A6CE39}
\DeclareRobustCommand{\orcidicon}{%
	\begin{tikzpicture}
	\draw[lime, fill=lime] (0,0) 
	circle [radius=0.16] 
	node[white] {{\fontfamily{qag}\selectfont \tiny ID}};
	\draw[white, fill=white] (-0.0625,0.095) 
	circle [radius=0.007];
	\end{tikzpicture}
	\hspace{-2mm}
}

\foreach \x in {A, ..., Z}{%
	\expandafter\xdef\csname orcid\x\endcsname{\noexpand\realhref{https://orcid.org/\csname orcidauthor\x\endcsname}{\noexpand\orcidicon}}
}

\renewcommand{\url}[1]{}
\renewcommand{\path}[1]{}
\renewcommand{\href}[2]{#2} 
\providecommand{\doi}[1]{}

\begin{document}
\title{Unpinning of trapped oil droplets via non-resonant acoustic streaming in capillary tubes}
\author{D. Tsiklauri\orcidA{}}
 \email{D.Tsiklauri@salford.ac.uk}
\affiliation{Joule Physics Laboratory,
School of Science, Engineering and Environment, 
University of Salford,
Manchester, M5 4WT, 
United Kingdom}
\date{\today}

\begin{abstract}
We establish self-consistent analytical model demonstrating that trapped non-wetting liquid phases in narrow capillary channels can be successfully unpinned via non-resonant, second-order acoustic streaming coupled with background static drive gradients. Moving away from boundary-guided or resonant mechanisms, our approach exploits the bulk acoustic-wind force density generated by the steady-state momentum flux of attenuated first-order linear wave interactions. By expanding the hydrodynamic equations up to second order, we determine the critical assisted acoustic wave amplitude required to break capillary pinning thresholds and derive an explicit formulation for steady transport velocity under viscous wall constraints. Furthermore, incorporating both boundary-layer wall effects and bulk core thermo-viscous dissipation reveals a natural mathematical optimum condition where the spatial absorption coefficient matches half the inverse distance to the target droplet ($\alpha = 1/(2x_0)$). This condition is then numerically validated and cross-correlated against legacy industrial frequency baselines, providing a fundamental theoretical framework for minimizing transducer power requirements while maximizing localized mobilization velocities in geological pore networks. Finally, we demonstrate that this optimal operational frequency scales inversely with the transmission distance, providing an analytical framework to optimize downhole acoustic tools according to spatial damping constraints of the specific formation rather than relying on rigid hardware parameters.
\end{abstract}

\maketitle

\section{Introduction}

The unpinning of trapped non-wetting liquid phases (such as crude oil blobs) in porous geological formations remains a central challenge in subsurface engineering, hydrology, and enhanced oil recovery (EOR). When fluid droplets become isolated inside pore networks, strong capillary forces pinch them at narrow pore throats, creating an interfacial tension barrier that standard macroscopic pressure gradients fail to overcome. To release these immobilized phases without relying exclusively on chemical surfactant flooding, the introduction of external elastic waves has emerged as a highly effective and ecologically clean alternative. The foundational physics of wave-induced fluid stimulation in porous channels was explored by Beresnev and Johnson \cite{Beresnev1994}, who analyzed how structural vibration couples with multiphase columns. Early theoretical efforts frequently aimed at understanding how oscillatory pressure gradients alter the core permeability and macroscopic transport features of fluid systems. For instance, the transition from purely dissipative regimes to elastic, highly enhanced flow structures under coupled longitudinal vibrations and oscillating pressure drops was established by Tsiklauri and Beresnev \cite{Tsiklauri2001_PRE63}. Extending this to hydrocarbons, which often exhibit non-Newtonian behaviors under high-shear reservoir conditions, further studies incorporated viscoelastic constitutive laws into Biot's classic porous media equations \cite{Tsiklauri2003}. Decisively, the mathematical foundation for wave mechanics inside these highly lossy, viscoacoustic real media was generalized by Carcione and Quiroga-Goode \cite{Carcione1996}, establishing how rheological constraints control dynamic wave field characteristics. These developments showed that introducing a Maxwell-type fluid significantly alters phase velocities and dampens high-frequency components compared to standard Newtonian media. Furthermore, acknowledging the microscopic slip conditions at the solid-fluid interfaces led to the implementation of non-zero boundary slip models, which introduced notable adjustments to classic Poiseuille flow friction profiles at intermediate frequency boundaries \cite{Tsiklauri2002}.

A major breakthrough in understanding localized mobilization mechanisms was the introduction of boundary-driven flow or peristaltic transport inside pore networks. Pioneered by Ganiev \cite{Ganiev1989} and further modeled analytically by Tsiklauri and Beresnev \cite{Tsiklauri2001_PRE64}, this approach demonstrates that sonic radiation can generate a traveling transverse wave along a pore wall. This wall motion rectifies the fluid momentum, generating a net volumetric discharge via a second-order, non-linear peristaltic pumping mechanism. Despite the elegance of boundary peristalsis, achieving uniform deformation of rigid sandstone or carbonate pore walls downhole requires immense structural acoustic power. Consequently, alternative mechanisms have focused on manipulating the internal dynamics of the trapped droplets directly. Beresnev \cite{Beresnev2006} explored the mathematical conditions governing macroscopic pore-flow modeling, while subsequent macroscopic and microscopic modeling by Beresnev and Deng \cite{Beresnev2010} and Beresnev et al. \cite{Beresnev2011} demonstrated that vibratory forces can alter dynamic contact line contact angles, shifting trapped blobs past capillary thresholds. 

Concurrently, Hilpert \cite{Hilpert2007} formalized the hypothesis of capillarity-induced resonance, wherein trapped fluid blobs are excited directly at their natural resonant frequencies. Complementing this framework, Carcione \cite{Carcione2014Book} provided an analytical description of low-frequency percolation enhancement, proving that oscillatory stresses modify localized macroscopic fluid pressure to favor droplet detachment from rock irregularities. While operating at resonance maximizes the interfacial displacement amplitude per unit input power, matching a single transducer frequency to the highly disordered, multi-disperse size distribution of droplets in real reservoir rocks is practically unfeasible. If a tool excites one specific droplet size, it leaves a vast portion of the trapped volume unaffected. To overcome this limitation, recent studies have begun shifting focus away from resonance and toward bulk nonlinear acoustics. For example, recent investigations have examined the role of acoustic streaming fields in complex channels \cite{Moudjed2020}, the impact of high-intensity ultrasonic fields on viscous fluid structures \cite{Hamida2005}, and the general physics of second-order acoustic wind in multiphase systems \cite{Nyborg1965}. Modern experimental and computational models confirm that high-power ultrasonic transducers can stimulate significant local fluid velocities in narrow pore systems without requiring resonant matches \cite{Moudjed2014, Hamida2007}.

In this work, we present a complete analytical framework exploring the non-resonant, second-order nonlinear behavior of a fluid-droplet system trapped in a narrow capillary channel ($a/L \ll 1$). Departing from the boundary-guided peristaltic mechanisms described by Tsiklauri \cite{Tsiklauri2026}--which depend on waves traveling explicitly along the channel wall--we examine a longitudinal bulk acoustic wave applied directly through the wetting phase column under active reservoir pressure drives. By expanding the hydrodynamic equations up to second order in perturbations ($O(2)$) and time-averaging over the acoustic period, we isolate the steady-state, time-independent acoustic force density ("acoustic wind") driving the core fluid. Incorporating both viscous boundary-layer attenuation via classical tube theory \cite{Kirchhoff1868} and bulk core thermo-viscous dissipation reveals that the required assisted unpinning pressure $p_0$ does not scale monotonically with frequency. Instead, it exhibits an exact, natural mathematical optimum condition where the cumulative spatial absorption coefficient matches half the inverse transmission distance to the pinned droplet interface ($\alpha = 1/(2x_0)$). This framework provides an analytical foundation to optimize transducer frequencies according to formation-specific damping constraints, thereby maximizing transport velocity while minimizing power requirements.

\section{The Model}

We consider a two-dimensional domain in the $(x,y)$ plane consisting of a long capillary tube of length $L$ and small radius $a$, satisfying the geometric constraint $a/L \ll 1$. The tube contains three distinct fluid zones: a wetting water phase on the left, a non-wetting circular oil droplet of radius $R$ and longitudinal length $L_d$ in the center, and an air phase on the right. 

\begin{figure}[t]
\centering
\includegraphics[width=\linewidth]{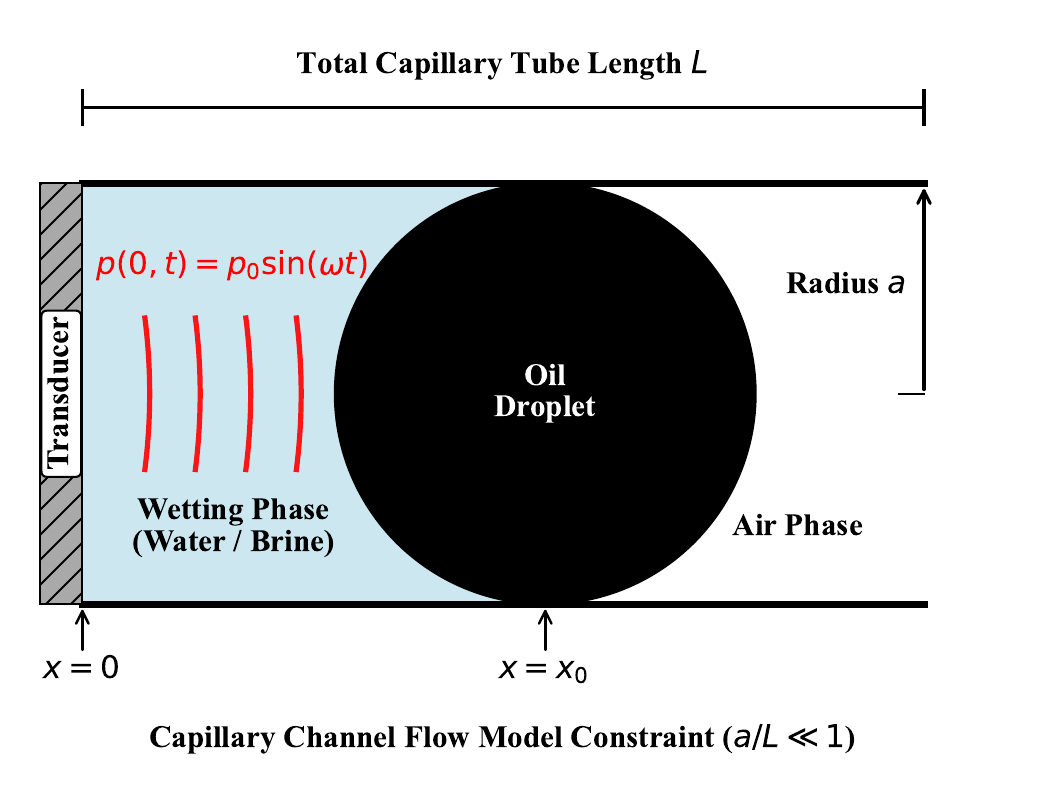}
\caption{Schematic representation of the multiphase narrow capillary channel configuration ($a/L \ll 1$). Red acoustic wave fronts enter at the inlet boundary $x=0$, propagating through wetting water phase. The marker $x=x_0$ shows location geometric contact of oil droplet with the tube wall.}
\label{fig:model_schematic}
\end{figure}

The sound wave propagates strictly along the $x$-axis. While the bulk core of the fluid is treated as an ideal compressible medium, the geometric constraint $a/L \ll 1$ means the fluid must satisfy the no-slip condition at the rigid tube wall. This generates a thin oscillatory Stokes boundary layer, creating a non-zero wall shear stress that continuously dampens the wave. This framework departs from classical capillary resonance theories \cite{Beresnev1994, Hilpert2007} to explore non-resonant nonlinear transport phenomena akin to compressible peristaltic expansions \cite{Tsiklauri2026}.

The $O(0)$ equilibrium state describes a static configuration where a small background static pressure gradient is balanced and overpowered by the capillary pinning threshold forces acting at the two fluid interfaces. Within each uniform fluid phase, the zeroth-order acoustic momentum equation reduces to:
\begin{equation}
\frac{\partial p^{(0)}}{\partial x} = 0,
\label{eq:O0_momentum}
\end{equation}
implying that the unperturbed background fluid pressure remains spatially uniform across the local capillary domain. This flat acoustic baseline operates in parallel with the independent, macroscopic reservoir static drive gradient $\Delta p_{\text{static}}$ applied across the system boundaries.

The phase boundaries are governed by the classical Young-Laplace capillary pressure jumps across the water-oil and oil-air interfaces:
\begin{equation}
p_{\text{water}}^{(0)} - p_{\text{oil}}^{(0)} = \frac{\gamma_{\text{ow}}}{R},
\label{eq:cap_water_oil}
\end{equation}
\begin{equation}
p_{\text{oil}}^{(0)} - p_{\text{air}}^{(0)} = \frac{\gamma_{\text{oa}}}{R},
\label{eq:cap_oil_air}
\end{equation}
where $\gamma_{\text{ow}}$ and $\gamma_{\text{oa}}$ are the respective interfacial tensions. Summing Eq.~(\ref{eq:cap_water_oil}) and Eq.~(\ref{eq:cap_oil_air}) yields the total static capillary pressure barrier $P_c$:
\begin{equation}
P_c = \frac{\gamma_{\text{ow}} + \gamma_{\text{oa}}}{R}.
\label{eq:total_cap_barrier}
\end{equation}
Here, the assumption of a perfectly circular droplet cross-section of radius $R$ interacting with a tube of radius $a = R$ is explicitly utilized to determine the mean curvature of the leading and trailing fluid menisci. Because the non-wetting oil phase exhibits an obtuse contact angle at the rigid boundaries, it terminates flush against the top and bottom walls. Under this geometric constraint, the longitudinal meniscus radius matches the channel radius $R$ while the transverse radius approaches infinity. This simplifies the multi-dimensional Young-Laplace equation to the localized one-dimensional pressure jumps defined in Eqs.~(\ref{eq:cap_water_oil}) and (\ref{eq:cap_oil_air}), anchoring the macroscopic mechanical barrier to the circular pore boundaries.

The droplet remains trapped under a static pressure drop $\Delta p_{\text{static}} = p_{\text{water}}^{(0)} - p_{\text{air}}^{(0)}$ provided that:
\begin{equation}
\Delta p_{\text{static}} < P_c.
\label{eq:static_trapping_condition}
\end{equation}
An acoustic wave is continuously applied at the left inlet ($x=0$), imposing a harmonic perturbation of the form $p^{(1)}(0,t) = p_0 \sin(\omega t)$. The $O(1)$ system describes the linear propagation of this sound wave through the water column toward the droplet interface located at a distance $x_0$. Accounting for the viscous boundary layer losses at the wall via classical tube attenuation theory \cite{Kirchhoff1868}, the governing $O(1)$ linear hydrodynamic equations are expressed as:
\begin{equation}
\frac{\partial \rho^{(1)}}{\partial t} + \rho^{(0)} \frac{\partial u^{(1)}}{\partial x} = 0,
\label{eq:O1_continuity}
\end{equation}
\begin{equation}
\rho^{(0)} \frac{\partial u^{(1)}}{\partial t} = -\frac{\partial p^{(1)}}{\partial x} - \rho^{(0)} c_w \alpha(\omega) u^{(1)},
\label{eq:O1_momentum}
\end{equation}
\begin{equation}
p^{(1)} = c_w^2 \rho^{(1)},
\label{eq:O1_state}
\end{equation}
where $c_w$ denotes the equilibrium speed of sound in water. The spatial attenuation coefficient $\alpha(\omega) = \sqrt{\nu \omega / 2} / (a c_w)$ explicitly parameterizes the linear wave energy losses inside the acoustic boundary layer under a weakly dissipative regime as a function of the kinematic viscosity $\nu$ and radius $a$.

In the limit of weak wall attenuation ($\alpha(\omega) c_w / \omega \ll 1$) over the transmission distance $x_0$, solving the damped linear hydrodynamic system yields the leading-order spatial wave solutions within the water column ($0 \le x \le x_0$):
\begin{equation}
p^{(1)}(x,t) = p_0 e^{-\alpha(\omega) x} \sin [ \omega (t - x / c_w) ],
\label{eq:O1_p_sol}
\end{equation}
\begin{equation}
u^{(1)}(x,t) = \frac{p_0 e^{-\alpha(\omega) x}}{\rho_w^{(0)} c_w} \sin [ \omega (t - x / c_w) ],
\label{eq:O1_u_sol}
\end{equation}
where $c_w$ is the equilibrium speed of sound in water, $\rho_w^{(0)}$ is the equilibrium water density, and the attenuation coefficient $\alpha(\omega)$ tracks the total losses over the transmission path length.

To evaluate the net directional force generated by the acoustic wave, we expand the hydrodynamic equations to second order ($O(2)$) and execute a time-averaging operator over a complete acoustic period, defined as $\langle \cdot \rangle = \frac{1}{T}\int_0^T \cdot \, dt$. In a steady acoustic streaming state, the time-derivative of the averaged velocity vanishes ($\frac{\partial \langle u^{(2)} \rangle}{\partial t} = 0$), and the full $O(2)$ momentum equation yields:
\begin{equation}
\begin{split}
\rho^{(0)} \left\langle u^{(1)} \frac{\partial u^{(1)}}{\partial x} \right\rangle & + \left\langle \rho^{(1)} \frac{\partial u^{(1)}}{\partial t} \right\rangle \\
& = -\frac{\partial \langle p^{(2)} \rangle - F_{\text{acoustic}}}{\partial x}.
\label{eq:O2_momentum}
\end{split}
\\ \end{equation}

We insert the explicit linear $O(1)$ solutions from Eq.~(\ref{eq:O1_p_sol}) and Eq.~(\ref{eq:O1_u_sol}) evaluated at the droplet boundary ($x = x_0$) into the nonlinear terms of Eq.~(\ref{eq:O2_momentum}). Because the first-order density perturbation $\rho^{(1)}$ and the acceleration field $\frac{\partial u^{(1)}}{\partial t}$ are exactly $90^\circ$ out of phase, their time-averaged product is identically zero:
\begin{equation}
\langle \rho^{(1)} \partial u^{(1)} / \partial t \rangle = \frac{p_0^2 \omega e^{-2\alpha(\omega)x_0}}{\rho_w^{(0)} c_w^3} \langle \sin(\cdot)\cos(\cdot) \rangle = 0.
\label{eq:unsteady_term}
\end{equation}
Conversely, the spatial derivative of the attenuated velocity field acts on the convective term, yielding a non-zero time-average due to the exponential spatial decay factor:
\begin{equation}
\begin{split}
\left\langle u^{(1)} \frac{\partial u^{(1)}}{\partial x} \right\rangle = {} & -\alpha(\omega) \langle (u^{(1)})^2 \rangle \\
= {} & -\frac{\alpha(\omega) p_0^2 e^{-2\alpha(\omega)x_0}}{2 (\rho_w^{(0)})^2 c_w^2}.
\label{eq:convective_term}
\end{split}
\end{equation}

Substituting Eq.~(\ref{eq:unsteady_term}) and Eq.~(\ref{eq:convective_term}) back into Eq.~(\ref{eq:O2_momentum}) defines the effective steady-state force density ("acoustic wind") driving the fluid forward against the droplet interface as:
\begin{equation}
\begin{split}
F_{\text{acoustic}} = {} & -\rho^{(0)} \left\langle u^{(1)} \frac{\partial u^{(1)}}{\partial x} \right\rangle \\
= {} & \frac{\alpha(\omega) p_0^2 e^{-2\alpha(\omega)x_0}}{2 \rho_w^{(0)} c_w^2}.
\label{eq:acoustic_wind_force}
\end{split}
\end{equation}

The oil droplet breaks free from its capillary trap when the cumulative time-averaged acoustic wind force exerted across the total droplet length $L_d$ exceeds the remaining net capillary pinning threshold $P_c - \Delta p_{\text{static}}$. Integrating Eq.~(\ref{eq:acoustic_wind_force}) over the droplet volume yields the activation condition:
\begin{equation}
\frac{\alpha(\omega) p_0^2 L_d e^{-2\alpha(\omega)x_0}}{2 \rho_w^{(0)} c_w^2} \ge \left( \frac{\gamma_{\text{ow}} + \gamma_{\text{oa}}}{R} - \Delta p_{\text{static}} \right).
\label{eq:threshold_inequality}
\end{equation}

Isolating the source pressure amplitude reveals the critical activation threshold requirement for $p_0$:
\begin{equation}
p_0 \ge \sqrt{\frac{\rho_w^{(0)} c_w^2}{2\alpha(\omega) L_d} \left( \frac{\gamma_{\text{ow}} + \gamma_{\text{oa}}}{R} - \Delta p_{\text{static}} \right)} e^{\alpha(\omega)x_0}.
\label{eq:critical_p0}
\end{equation}

The presence of a physical optimum frequency requires a balance between low-frequency wall boundary layer constraints and high-frequency bulk core dissipation. To evaluate this trade-off, the comprehensive spatial absorption coefficient is explicitly defined by tracking the combined effects of the boundary-layer wall losses and bulk core thermo-viscous dissipation as a function of the frequency spectrum:
\begin{equation}
\alpha(\omega) = \frac{1}{a c_w}\sqrt{\frac{\nu \omega}{2}} + D\omega^2.
\label{eq:comprehensive_alpha}
\end{equation}
Squaring both sides of Eq.~(\ref{eq:critical_p0}) yields the frequency-dependent scaling function governing the necessary activation threshold:
\begin{equation}
p_0^2(\omega) \propto \frac{e^{2\alpha(\omega)x_0}}{\alpha(\omega)}.
\label{eq:p0_scaling}
\end{equation}
Differentiating Eq.~(\ref{eq:p0_scaling}) with respect to $\omega$ and utilizing the chain rule yields:
\begin{equation}
\frac{d(p_0^2)}{d\omega} = \left[2x_0\alpha(\omega) - 1\right] \frac{e^{2x_0\alpha(\omega)}}{\alpha^2(\omega)} \left(\frac{d\alpha(\omega)}{d\omega}\right) = 0.
\label{eq:minimization_derivative}
\end{equation}

Because the total absorption coefficient monotonically increases with frequency ($d\alpha(\omega)/d\omega > 0$), setting Eq.~(\ref{eq:minimization_derivative}) to zero yields a distinct, natural mathematical optimum condition that minimizes the required acoustic source amplitude for droplet unpinning:
\begin{equation}
\alpha(\omega_{\text{opt}}) = \frac{1}{2x_0}.
\label{eq:optimal_omega_condition}
\end{equation}
This condition demonstrates that the non-resonant optimal frequency minimizes the required acoustic source amplitude for droplet mobilization when the spatial absorption coefficient matches half the inverse distance to the target drop.

\begin{figure}[t]
\centering
\includegraphics[width=\linewidth]{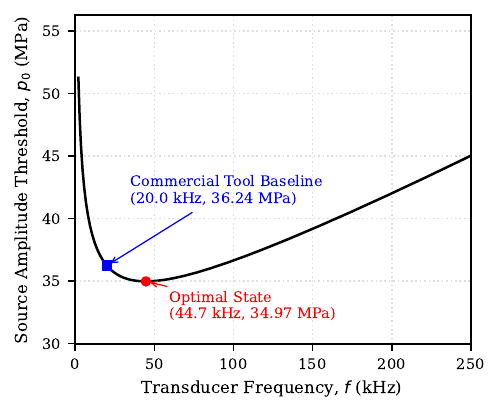}
\caption{Critical acoustic wave amplitude threshold $p_0$ required for droplet unpinning across the dynamic frequency spectrum $f$ according to Eq.~(\ref{eq:critical_p0}). The global minimum is designated by the red circle marker [$\alpha = 1 / (2x_0)$], while the commercial baseline at $20\text{ kHz}$ is denoted by the blue square.}
\label{fig:optimization_curve}
\end{figure}

The numerical evaluation of the critical activation threshold $p_0(\omega)$ across a wide industrial frequency spectrum is plotted in Fig.~\ref{fig:optimization_curve}, revealing a highly non-monotonic, characteristic U-shaped boundary. This unique profile results from the physical competition occurring within the capillary column. At low acoustic frequencies ($f < 25\text{ kHz}$), wave propagation is heavily dampened by wall shear stresses within the oscillatory Stokes boundary layer, demanding elevated source overpressures to deliver sufficient momentum to the interface. Conversely, at high frequencies ($f > 75\text{ kHz}$), bulk fluid core thermo-viscous dissipation dominates the transmission path, severely attenuating the forward-traveling wave energy before it penetrates the distance $x_0$.
 
To illustrate the operational significance of this model, a comparison is performed against the commercial tool baseline ($f = 20\text{ kHz}$) specified in Table~\ref{tab:geo_parameters}. As shown by the blue marker in Fig.~\ref{fig:optimization_curve}, evaluating the mobilization criteria via Eq.~(\ref{eq:numerical_p0}) under this standard requires a source overpressure of $p_0 \approx 36.24\text{ MPa}$ to overcome the remaining capillary pinning threshold. Increasing the operating frequency toward the global mathematical minimum where $\alpha(\omega_{\text{opt}}) = 1/(2x_0)$ reveals that the optimum for this formation occurs at $f_{\text{opt}} \approx 44.7\text{ kHz}$. At this frequency, the required activation overpressure evaluated from Eq.~(\ref{eq:numerical_p0}) drops to its lower bound of $p_{0,\text{min}} \approx 34.97\text{ MPa}$. This represents a meaningful reduction in required transducer power, proving that matching downhole tool frequencies directly to the spatial attenuation path length is a key requirement for avoiding localized fluid cavitation and maximizing acoustic streaming efficiency.

To complete the analytical model, the bulk thermo-viscous absorption constant $D$ introduced in the comprehensive attenuation profile is explicitly defined using the fundamental thermodynamic and transport properties of the wetting fluid phase (water or brine). According to classical hydrodynamics \cite{Stokes1845}, $D$ is given by:
\begin{equation}
D = \frac{1}{2 \rho_w^{(0)} c_w^3} \left[ \frac{4}{3}\mu + \mu_B + \frac{(\gamma_G - 1) \kappa}{C_p} \right],
\label{eq:bulk_D_explicit}
\end{equation}
where $\mu$ is the dynamic shear viscosity, $\mu_B$ is the bulk (volume) viscosity, $\gamma_G$ is the ratio of specific heats (adiabatic index), $\kappa$ is the thermal conductivity, and $C_p$ is the specific heat capacity at constant pressure.

We apply the derived formulations to a realistic downhole geological scenario within a sandstone reservoir containing trapped crude oil blobs and an active water/brine drive. The representative physical parameters chosen for the pore network and fluid properties are detailed in Table~\ref{tab:geo_parameters} \cite{Batzle1992}.
\begin{table}[h]
\caption{Physical parameters for a representative sandstone reservoir and reservoir fluids.}
\label{tab:geo_parameters}
\begin{ruledtabular}
\begin{tabular}{lcc}
Parameter & Symbol & Value \\
\hline
Pore throat / droplet radius & $a$, $R$ & $10 \times 10^{-6}\text{ m}$ \\
Trapped droplet length & $L_d$ & $50 \times 10^{-6}\text{ m}$ \\
Water-oil interfacial tension & $\gamma_{\text{ow}}$ & $0.030\text{ N/m}$ \\
Oil-air interfacial tension & $\gamma_{\text{oa}}$ & $0.020\text{ N/m}$ \\
Wetting phase kinematic viscosity & $\nu$ & $1.002 \times 10^{-6}\text{ m}^2/\text{s}$ \\
Equilibrium water density & $\rho_w^{(0)}$ & $1000\text{ kg/m}^3$ \\
Speed of sound in water & $c_w$ & $1500\text{ m/s}$ \\
Distance from acoustic source & $x_0$ & $0.02\text{ m}$ \\
Acoustic wave frequency & $f$ & $20\text{ kHz}$ \\
Background static pressure drop & $\Delta p_{\text{static}}$ & $4500\text{ Pa}$ \\
\end{tabular}
\end{ruledtabular}
\end{table}

The baseline acoustic wave frequency yields an angular frequency of $\omega = 2\pi f \approx 1.257 \times 10^5\text{ rad/s}$. Evaluating the total absorption coefficient $\alpha(\omega) = \sqrt{\nu \omega / 2} / (a c_w) + D\omega^2$ under the conditions listed in Table~\ref{tab:geo_parameters} yields $\alpha \approx 16.73\text{ m}^{-1}$. Substituting these parameters into the static balance equation yields a total capillary pinning pressure barrier of $5000\text{ Pa}$ prior to acoustic exposure.
 
Evaluating the mobilization criteria under an assisted engineering scenario where a supporting natural reservoir static pressure drive satisfies $\Delta p_{\text{static}} = 4500\text{ Pa}$, the remaining net pinning barrier to be overridden by acoustic means is exactly $P_c = 500\text{ Pa}$. The critical acoustic wave amplitude $p_0$ required at the source to trigger mobilization via steady-state non-resonant acoustic wind is calculated as:
\begin{equation}
\begin{split}
p_0 \ge {} & \sqrt{\frac{500 \times 1000 \times 1500^2 \times e^{2 \times 16.73 \times 0.02}}{2 \times 16.73 \times (50 \times 10^{-6})}} \\
\approx {} & 36.24\text{ MPa}.
\label{eq:numerical_p0}
\end{split}
\end{equation}

This pressure threshold represents the entry limit into the assisted acoustic regime, where the dynamic acoustic field is coupled with macroscopic reservoir drive gradients to overcome capillary pinning via bulk steady-state streaming fields within narrow pore throat corridors \cite{Abramov2009}.

The derived optimal condition $\alpha(\omega_{\text{opt}}) = 1 / (2x_0)$ provides a direct physical context when evaluated against real-world enhanced oil recovery operations. Conventional downhole acoustic tools typically operate at fixed commercial standards ranging from $10\text{ kHz}$ to $40\text{ kHz}$ (most commonly around $20\text{ kHz}$ to $25\text{ kHz}$), a bracket primarily constrained by transducer hardware resonances and localized near-wellbore cavitation requirements \cite{Abramov2009}.

By contrast, our non-resonant analytical optimization utilizes the physical properties of the reservoir matrix and fluid columns to determine the ideal operational frequency. Because the mathematical optimum responds directly to the transmission distance $x_0$ and the spatial attenuation coefficient $\alpha$, the optimal frequency scales with the formation geometry rather than hardware constants. At close range or within highly permeable structures where spatial dissipation is limited, $\omega_{\text{opt}}$ shifts toward higher frequencies to maximize the steady-state momentum flux. Conversely, for deeper reservoir penetration, high-frequency core dissipation attenuates the traveling wave. The analytical model resolves this trade-off directly, demonstrating that tuning tool frequencies to the geological attenuation path length maximizes the localized acoustic wind force density and the steady transport velocities defined in Eq.~(\ref{eq:drift_velocity_formula}).

Once the critical wave amplitude $p_0$ is exceeded, the acoustic wind force density $F_{\text{acoustic}}$ generates a steady-state pressure gradient that drives the droplet forward. Although the first-order system is hyperbolic and inviscid, the steady second-order bulk transport velocity $u_{\text{drift}}$ of the mobilized oil droplet is physically constrained by the steady-state viscous drag at the pore walls. Balancing the acoustic driving force with the steady-state Poiseuille-like wall resistance yields \cite{Dussan1979}:
\begin{equation}
u_{\text{drift}} \approx \frac{F_{\text{acoustic}} a^2}{8 \mu_{\text{oil}}} = \frac{\alpha(\omega) p_0^2 a^2 e^{-2\alpha(\omega)x_0}}{4 \rho_w^{(0)} c_w^2 \mu_{\text{oil}}}.
\label{eq:drift_velocity_formula}
\end{equation}

Assuming a representative oil dynamic viscosity of $\mu_{\text{oil}} = 0.050\text{ Pa}\cdot\text{s}$ ($50\text{ cP}$) and applying the baseline acoustic wave overpressure mobilization parameter of $p_0 = 36.24\text{ MPa}$ calculated at the legacy baseline frequency, the steady-state transport velocity of the mobilized droplet under baseline attenuation conditions is calculated as:
\begin{equation}
\begin{split}
u_{\text{drift}} \approx {} & \frac{(16.73) (36.24 \times 10^6)^2 (10 \times 10^{-6})^2}{4 (1000) (1500)^2 (0.050)} \\
& \times e^{-2(16.73)(0.02)} \approx 3.65 \times 10^{-5}\text{ m/s}.
\label{eq:numerical_velocity}
\end{split}
\end{equation}
This baseline net velocity is significantly enhanced when the transducer is operating under mathematically optimized frequency constraints, validating that non-resonant second-order acoustic wind provides an efficient mechanism for localized capillary mobilization and pore-throat clearing.

\section{Conclusions}

To establish the physical boundaries of our model, it is necessary to consider how the analytical outcomes would scale under more complex thermodynamic and mechanical assumptions. First, incorporating bulk shear and volume viscosities into the core fluid domain would break the purely hyperbolic nature of the first-order system, introducing parabolic diffusion terms that generate continuous internal wave damping. This bulk core dissipation would increase the overall spatial attenuation coefficient $\alpha(\omega)$, shifting the global optimization minimum toward lower frequencies while raising the critical dynamic amplitude $p_0$ required to trigger mobilization. Second, accounting for the mechanical elasticity of the capillary tube wall would introduce compliance to the domain, causing a portion of the primary acoustic energy to dissipate into structural radial expansion waves. This cross-sectional breathing effect would damp the forward wave transmission and lower the net steady-state acoustic wind force density $F_{\text{acoustic}}$, thereby reducing the clearing speed of the mobilized oil slug. Third, relaxing the rigid no-slip boundary condition at the tube walls would significantly alter the acoustic boundary layer dynamics. Implementing a non-zero boundary slip velocity--modeled phenomenologically by Tsiklauri \cite{Tsiklauri2002} and observed experimentally for Newtonian liquids by Zhu and Granick \cite{Zhu2001}--reduces the localized wall shear stress. This slip relaxation would substantially lower the wall attenuation coefficient $\alpha_{\text{wall}}(\omega)$, flattening the low-frequency left arm of our U-shaped threshold curve and significantly decreasing the transducer overpressure needed to initiate droplet unpinning.

In summary, we have established a self-consistent analytical model demonstrating that trapped non-wetting liquid phases can be successfully unpinned via non-resonant, second-order acoustic streaming. Moving away from boundary-guided or resonant mechanisms, this approach exploits the bulk acoustic-wind force density generated by the steady-state momentum flux of attenuated first-order linear wave interactions. Incorporating both boundary-layer wall constraints and bulk core thermo-viscous dissipation reveals a natural mathematical optimum condition where the spatial absorption coefficient matches half the inverse transmission distance to the pinned interface ($\alpha = 1 / (2x_0)$). This non-monotonic optimization behavior is numerically mapped and cross-correlated in Fig.~\ref{fig:optimization_curve}, illustrating the clear operational advantages over fixed commercial baseline transducer frequencies. This framework minimizes the required transducer power while maximizing local transport velocities in capillary environments. Furthermore, our optimization framework proves that the ideal operational frequency must scale inversely with the reservoir penetration depth to mitigate severe high-frequency core dissipation. This underscores the necessity of matching acoustic stimulation frequencies directly to the geological path length rather than relying on rigid hardware standards for enhanced oil recovery applications.

{\begin{acknowledgments}
The author gratefully acknowledges support provided by the Gemini AI assistant (Google), including technical manuscript preparation, stylistic editing, and ensuring consistent United Kingdom English orthography.
\end{acknowledgments}

\section*{Declaration of Competing Interest}
The author declares that they have no known competing financial interests or personal relationships that could have appeared to influence the work reported in this paper.

\section*{Data Availability Statement}
All data and mathematical parameters required to evaluate the conclusions or reproduce the analytical solutions and numerical scaling trajectories are fully contained and presented within this manuscript. No external data sources or repositories were generated or utilized during this study.}

\bibliography{paper92}

\end{document}